# More Papers, Not More Output: Collaboration and Publication Trends among Early-Career Researchers

Leibel, Christian

ORCID: 0009-0006-6893-4901

*christian.leibel.extern@gv.mpg.de (corresponding author)*

Science Policy and Strategy Department, Administrative Headquarters of the Max Planck Society, Germany

Department of Sociology, LMU Munich, Germany

# 1 Summary

Concerns about publication pressure often assume that early-career researchers now produce more papers than comparable researchers did in the past. However, publication counts may give a misleading impression of individual publication output because research has become increasingly collaborative. Researchers can list more papers on their CVs not only by producing more work themselves, but also by working with more coauthors and sharing credit for the same papers. Whether the collaboration-adjusted output of early-career researchers has increased over time has rarely been examined empirically. This study addresses this gap by analysing publication records from more than 200,000 early-career researchers at Western research institutions across five major science, engineering, and health fields between 1996 and 2024. Recent cohorts appear on more papers than earlier cohorts, but they also work in much larger research teams. Once publication credit is divided among coauthors, collaboration-adjusted output remains stable or declines across all fields. These findings show that research evaluation metrics that ignore collaboration are increasingly poorly aligned with how scientific work is produced. Rising publication counts alone should therefore not be taken as evidence that early-career researchers are producing more individual publication output.

# 2 Introduction

"Today I wouldn't get an academic job. It's as simple as that. I don't think I would be regarded as productive enough" (The Guardian, 6 Dec 2013). This quote by Nobel Prize-winning physicist Peter Higgs is counterfactual, but many researchers may find it compelling because it expresses a widely shared concern: that researchers are increasingly judged by how much they publish rather than by the scientific merit of their work. This concern is rooted in major changes to the academic labour market since the 1960s, when Higgs conducted his influential research. Over this period, the number of PhD holders has expanded much faster than the number of available tenure-track positions [1,2]. Competition for long-term academic employment has intensified, and average academic career lengths have declined as more researchers enter and quickly leave the system [3]. Critics of academia's contemporary publication culture argue that intense competition for permanent positions and short funding cycles create an environment in which only those who produce a constant stream of quantifiable output can hope for long-term career survival in academia [4].

Scholars worry about publication pressure not only because it may negatively affect researchers' wellbeing and mental health [5–7], but also because it may damage science by creating problematic incentives [8]. Researchers may favour projects that are likely to produce rapid and publishable results over more ambitious, time-consuming projects [4,9]. If researchers with long publication lists are more likely to secure fundings and permanent positions, while researchers producing fewer but high-quality studies are selected out of academia, then questionable research practices become increasingly competitive [10,11]. Prior research has described this possibility as the natural selection of bad science [12,13].

At first sight, publication trends seem to confirm this pessimistic account. Several studies report that recent cohorts of researchers publish more papers before tenure than earlier cohorts [2,14,15]. For example, a study on evolutionary biologists at France's National Centre for Scientific Research (Centre national de la recherche scientifique) found that junior researchers employed in 2015 had published almost twice as many articles compared to their counterparts who were employed in 2003. Such findings have been interpreted as evidence of "salami slicing," in which researchers divide scientific results into multiple, less substantial publications to artificially increase their publication counts [14].

However, this interpretation rests on an important assumption: that the number of papers on a researcher's CV is a valid measure of that researcher's individual publication output. This assumption is increasingly problematic because science has become more collaborative. Across all fields of research, papers are now more often produced by teams rather than by individual researchers [16–23].

Because each coauthor can claim the same publication, researchers in larger collaboration networks can share both the workload and the resulting publications with their coauthors. Rising publication counts may therefore reflect two different processes: an increase in individual publication output, or an increase in collaboration that allows more researchers to claim credit for the same papers.

Distinguishing between these explanations is essential for assessing claims about publication pressure and its consequences for research quality. This requires evidence on whether early-career researchers' individual publication output has changed after accounting for collaboration. Yet such evidence remains scarce. Although one study of leading U.S. sociology departments found rising publication rates even after accounting for collaboration [2], its findings may not generalize beyond this particular institutional and disciplinary setting. The only large-scale study reached the opposite conclusion, finding no growth in collaboration-adjusted publication output [19]. However, it examined researchers who began publishing in the twentieth century. Whether more recent cohorts have increased their individual publication output therefore remains an open empirical question.

This study addresses this gap by analysing publication records from more than 200,000 early-career researchers at Western research institutions across five fields: chemistry and materials sciences, earth and environmental sciences, engineering, health sciences, and life sciences. It examines whether early-career researchers are increasing their individual publication output or instead appearing on more papers because research is increasingly conducted in teams. The main analysis covers researchers who started publishing between 1996 and 2015 and supporting analyses extend to cohorts up to 2020. Across these analyses, recent cohorts show no robust increase in collaboration-adjusted publication output. The results show that publication counts should be understood alongside the growing importance of teamwork in contemporary research. The study concludes by discussing the implications for research policy and research evaluation.

# 3 Background

### 3.1 The natural selection of bad science

A central concern in debates about "publish or perish" is that academic careers reward quantity at the expense of quality. Numerous opinion papers [10,11,24,25] and empirical studies [8,26–28] argue that researchers facing strong publication pressure may have incentives to produce more papers, even when doing so weakens the rigor or scientific value of their work. In this view, scientific output resembles a zero-sum game: more papers mean less time and care devoted to each one.

This logic is stated most explicitly in the formal model of the "natural selection of bad science" [8], hereafter NSBS. The model was developed to explain why low statistical power and other weak research practices persist in fields such as psychology, despite longstanding awareness of these problems and repeated calls for improvement [29,30]. The NSBS model is useful for the present study because it makes clear how publication counts enter arguments about publication pressure and the quality of science.

NSBS describes academia as a competitive system in which researchers are rewarded mainly for producing a constant stream of quantifiable output. Researchers and research groups with longer publication lists are more likely to receive funding, remain in academia, and influence the next generation of researchers. Because publication output is strongly rewarded, practices that increase the number of publishable papers spread, even when they do not improve the reliability or scientific value of research.

The problem, according to NSBS, is that producing many papers conflicts with practices that require time but support research quality. Recruiting sufficiently large samples, checking data and analyses, and ensuring that findings are reproducible all require time and effort. Under the NSBS model, researchers who invest more time in these activities produce fewer papers and are therefore less competitive than researchers who produce papers more quickly. In the long term, this selection mechanism produces two outcomes: Researchers who manage to remain in academia publish more papers, while weaker research practices and lower-quality research become more common.

NSBS's explanation of why poor research practices persist rests on two assumptions: that researchers are publishing more papers over time, and that producing more papers requires a reduction in the scientific merit of each paper. As I show below, the growth of research teams changes how both assumptions should be understood. Researchers can increase the number of papers on their CVs without necessarily reducing the quality of their work. They can do so by collaborating more.

### 3.2 Why collaboration matters

The number of papers that researchers can list on their CVs depends on three variables: the number of distinct papers produced ($P$), the number of researchers ($N$), and the average number of authors per paper ($k$). If every author lists each coauthored paper on their CV, the total number of publication entries across the population is:

$$\text{Total CV entries: } P \times k$$

Assuming, for the sake of simplicity, that authorship is distributed evenly across researchers, then the average number of papers listed on an individual researcher's CV is:

$$\text{Average papers per CV: } (P \times k)/N$$

In the formulas above, $P$ can be interpreted as the real publication output of $N$ researchers. Because the number of CV entries depends on the product $P \times k$, increasing collaboration among researchers can increase publication lists even when the number of published papers remains unchanged. This mechanism matters because collaboration has increased in recent decades. It is well documented in the literature that the average number of authors per paper has grown substantially over time across all research fields [16–23]. Contemporary researchers can therefore have longer publication lists partly because they work in larger teams, not necessarily because their total publication output (i.e. $P$) has increased.

The original NSBS model does not incorporate the growth of research teams. Yet, collaboration can easily be understood within the broader incentive logic of NSBS. If academic careers reward long publication lists, then collaboration is indirectly rewarded because it allows researchers to appear on more papers. In some contexts, collaboration is also encouraged directly. Many countries use performance-based evaluation systems to monitor and steer the output of research institutions [31]. Such evaluation systems typically reward collaboration and especially international collaboration. In Germany, for example, some research institutions are required to report their output to the federal government, and international collaboration is an important element of these reports [32].

Of course, researchers do not collaborate only in response to external incentives. Collaboration may also increase because research itself has become more complex and demanding. For example, a historical analysis of life-science papers found that articles published in 2013 contained more figures and substantially more supplementary material than articles published in the early 1990s [23]. Growing collaboration may therefore reflect the increasing amount and variety of work required to produce a publishable study. Lastly, collaboration has likely increased in part because modern communication technologies have made it much easier for researchers to cooperate, especially across long distances.

Regardless of why researchers collaborate more, this trend changes the meaning of a growing publication record. Collaboration is not inherently a sign of poor research practice. Longer publication lists alone therefore do not show that researchers are producing more papers by cutting corners. Before drawing that conclusion, we need to ask a different question: do researchers publish more output than earlier cohorts once the growth of collaboration is taken into account?

### 3.3 Results and limitations of prior research

The ideal collaboration-adjusted measure of individual publication output would be $P/N$, i.e. the number of distinct papers per distinct researcher. Determining distinct publications $P$ is comparatively straightforward because papers typically contain persistent identifiers, such as DOIs, in their metadata. However, because no universal, comprehensive, and historically stable identifier exists for researchers, it is currently not possible to determine $N$. This fundamental methodological challenge is likely the main reason why empirical evidence on on historical trends in individual publication rates is extremely scarce.

The alternative to the population-level measure $P/N$ is to construct an author-level metric that adjusts for collaboration by assigning contribution shares to each author of a multi-author publication. A researcher's publication output is then defined as the sum of their contribution shares across all their papers. Such a collaboration-adjusted measure of individual contribution shares is the central outcome examined in this study.

Methods that assign each author only a fraction of the credit for a multi-author publication are known in scientometrics as fractional counting methods. To my knowledge, only one study has used fractional counting to distinguish changes in individual publication output from changes in academic collaboration [19]. The study conducted a large-scale analysis of researchers whose publication careers began between 1900 and 1998. It found no increase in fractional publication output over the twentieth century. Even among more recent cohorts who began publishing after 1980, the study found no evidence that collaboration-adjusted output had increased.

This earlier study provides important evidence that the growth of early-career researchers' publication records may be driven by intensifying collaboration rather than by increasing individual output. However, two questions remain unresolved. First, it is unclear whether this conclusion applies to contemporary cohorts. The study examined researchers who began publishing in the twentieth century, whereas the most rapid expansion of scholarly publishing has occurred since then. Academic publishing has continued to grow, and current estimates suggest that the global body of scientific literature now doubles in size approximately every 17 years [33]. The transition from print to digital publishing made it easier and faster to disseminate research, and the COVID-19 pandemic further accelerated publication processes in many biomedical journals [34]. These developments make it necessary to reassess whether early-career researchers' collaboration-adjusted output has remained stable in more recent periods.

Second, it remains unclear how sensitive the earlier findings are to the data source and analytical choices. Current evidence relies on the Web of Science. Because bibliographic databases differ in their curation, classification, and coverage of documents, trends identified in one database may not be reproduced in another [35]. A more fundamental limitation concerns the study's counting method. It used equal fractional counting, which assigns every coauthor the same share of a paper. For example, each author of a five-author paper receives 0.2 publication credits. However, this approach is not supported by the current literature on author contributions, which consistently shows that contributions to research papers are almost always unequal [36–45]. Papers typically include leading contributors, who conduct a large share of the intellectual and practical work, and supporting contributors, who make smaller or more specialized contributions. This is especially relevant for early-career researchers, who in many cases make more than equal contributions to their papers because they are expected to demonstrate their ability to conduct research independently [36]. Equal fractional counting may therefore provide an incomplete measure of their individual output.

This study addresses the gap in the current literature by examining historical trends in four aspects of publication output: (1) the number of papers in researchers' publication records; (2) the number of authors per paper; (3) the average share of the workload that early-career researchers contribute to their papers; and (4) their total collaboration-adjusted publication output. A secondary goal is to assess whether the findings are consistent across bibliographic databases and empirically validated fractional-counting methods. The study therefore not only updates earlier evidence, but also examines whether conclusions about individual publication output are robust to data-source and measurement choices.

# 4 Data and Methods

## 4.1 Bibliometric metadata

All data used in this study were collected from the data infrastructure of the German Kompetenznetzwerk Bibliometrie. The Kompetenznetzwerk Bibliometrie is a national consortium of 22 research institutions that hosts bibliometric data from several providers, including Scopus (Elsevier) and OpenAlex. The Kompetenznetzwerk Bibliometrie is a valuable source of bibliometric metadata because it is curated by experts who apply additional quality-assurance procedures beyond those conducted by the original data providers [46]. I used the latest database snapshots available at the time of analysis, dating from January 2026. Some records from 2025 may be missing because of indexing delays. To account for this possibility, I tracked researchers' publication activity only through 2024.

## 4.2 Construction of individual publication records

The process of reconstructing individual publication records from bibliometric metadata is known as author disambiguation. It uses information such as author names, institutional affiliations, coauthor networks, and research topics to determine which papers were written by the same person. Ideally, author-disambiguation methods would be evaluated against a large-scale ground-truth dataset, that is, a dataset in which the true author of every publication is known with certainty. Because no such comprehensive dataset exists, it is not possible to determine directly, for all researchers and publications in a large bibliometric database, whether an algorithm has assigned each paper to the correct person.

In the absence of large-scale ground truth, author disambiguation requires triangulating publication records across several data providers (see Appendix, Section S1, for supplementary information on author disambiguation methods). Rather than relying on one author-disambiguation procedure and one source of bibliometric data, this study assesses whether multiple providers, each with distinct data sources, matching procedures, and potential errors, agree that a publication was authored by a given individual. Agreement across these independent sources increases confidence that the observed patterns reflect a researcher's actual publication behaviour.

The author identifiers provided by OpenAlex and Scopus form the foundation of the triangulation strategy used in this study. These identifiers are available at large scale because they are algorithmically generated. Scopus author identifiers have long been used in research evaluation and have shown satisfactory accuracy in prior research [47]. I first matched publications in OpenAlex and Scopus using DOIs and retained only publications that corresponded to exactly one record in each database. This produced a deduplicated set of publications indexed in both sources. I then extracted all author identifiers attached to these publications, creating a dataset of publication–author pairings.

For each matched author, I constructed three publication records: the full record in Scopus, the full record in OpenAlex, and a combined high-confidence record consisting only of DOIs assigned to the same individual in both databases. The combined record is central to the study's measurement strategy. It contains publications for which two independent databases, using different metadata and different author-disambiguation algorithms, agree that the same individual authored the same paper.

I applied three quality-assurance filters before retaining an author profile. First, the OpenAlex and Scopus records had to assign the author the same first publication year, because this year determines cohort membership. Second, I retained only unambiguous one-to-one matches between author identifiers: publications assigned to a given Scopus author ID had to correspond to exactly one OpenAlex author ID, and vice versa. Third, the combined high-confidence record had to cover at least 90% of the full publication record in both OpenAlex and Scopus or deviate by no more than 1 paper. In other words, nearly all papers assigned to an author in either database also had to be assigned to the same author in the other database. This strict overlap criterion ensures that the analyses are based on publication records for which independent disambiguation systems strongly agree.

### 4.3 Data collection and processing

All publication output metrics are based on papers, defined as articles, reviews, and conference papers. Conference papers are included because they are an important publication channel in some fields, especially engineering [48]. Researchers are assigned to cohorts based on the publication year of their first observed paper. Prior research has shown that first publication year is a good proxy for academic age, especially in the fields analysed in this study [49]. The analysis begins with the 1996 cohort, the earliest year for which the Kompetenznetzwerk Bibliometrie has licensed Scopus data.

Because academic careers vary widely in length, publication output is measured within a fixed early-career window: the first ten years of publication activity. The target population therefore consists of researchers whose publication activity lasted at least ten years. This threshold intentionally excludes individuals who publish only a small number of papers, for example while completing a PhD, and then leave academia shortly afterward. These individuals are outside the study's target population because arguments about publication pressure, competition for resources, and tenure positions apply primarily to researchers who pursue sustained academic careers. I conducted a supporting analysis using a broader population of early-career researchers with at least five years of publication activity (section 5.3).

Large-scale bibliometric analysis involves a trade-off between sample size and measurement validity. Because the findings may inform debates about research evaluation and science policy, I prioritized metadata quality over coverage. I therefore excluded observations for which bibliometric metadata are likely to be insufficiently reliable for reconstructing individual publication records or estimating individual publication output. In the following, I focus on the substantive logic of the data processing; a detailed technical description is provided in the Appendix (Section S2)

First, I restricted the analysis geographically. Author-disambiguation methods rely partly on researcher names, and name-based identification is less reliable in settings where many names become highly similar after transliteration into the Latin alphabet. This problem is especially relevant for some world regions, including parts of Southeast Asia. To reduce such disambiguation errors, I restricted the analyses to authors primarily affiliated with institutions in North America, the EU-15, Australia, and New Zealand. This regional restriction follows prior research on individual publication rates [19].

Second, I excluded authors who typically publish in very large teams. This restriction prevents population-level averages from being distorted by hyperauthored papers with exceptionally large numbers of contributors [50,51]. It is also substantively justified because such papers often come from highly specific research contexts, such as CERN high-energy particle physics experiments, which are not directly comparable to standard research settings [52]. Moreover, fractional counting methods may be difficult to interpret in extremely large collaborations [44,45]. Researchers were retained only if they had no more than 20 coauthors on average, following a threshold recommended in prior research [50].

Third, I restricted the field coverage. Bibliometric databases primarily index peer-reviewed journal articles in international journals. Applying bibliometric methods to fields without article-centred publication cultures, as is often the case in the social sciences and humanities [2,53], risks misrepresenting publication dynamics. The analysis therefore focuses on science, engineering, and health fields, where journal articles and conference papers are dominant forms of scholarly communication [48]. I excluded physics and mathematics because they have distinctive authorship and publication practices, including very large collaborations in parts of physics and frequent alphabetical author ordering in both fields. These features make estimates of individual publication output less comparable to those for the other fields analysed here [48].

## 4.4 Counting method

This study utilizes the A-index counting method [54]. The A-index rests on three axioms. Below, $x_i$ denotes the share of total contribution attributed to the author in position $i$ and $n$ denotes the number of authors:

1) The contributions of all authors sum to 1: $\sum_i^n x_i = 1$;
2) Contributions decline with author position: $x_{i-1} > x_i$;
3) All distributions of author contributions that satisfy the first two assumptions are treated as equally likely.

The A-index assumes that the order in which authors are listed carries information about their contributions. The first author is treated as the main contributor; the second author receives a smaller share of the total contribution than the first author but a larger share than the third author; and so forth. In other words, axiom 2 defines an upper and lower bound for an author's contribution based on their position in the author list. Axiom 3 "essentially postulates every scenario is equally possible between a coauthor doing his/her very best or the bare minimum" [54]. The contribution share assigned to the author in position $i$ is then defined as the expected value of $x_i$ over all permissible scenarios.

$$E(x_i) = \frac{1}{n}\sum_{j=i}^{n}\frac{1}{j}$$

The A-index framework is consistent with current best practice in research on the measurement of author contributions. Its main strength is that it recognizes that individual contributions cannot be observed directly in publication metadata. Other counting methods address this limitation by applying more or less arbitrary weighting schemes. By contrast, the A-index makes this uncertainty explicit and addresses it by modeling author contributions using a theoretically grounded probability distribution.

The validity of the A-index is supported by empirical evidence. First, the alternative convention of listing authors alphabetically rather than in order of contribution has become rare in most fields and is mainly practiced in fields excluded from the analysis [48]. Second, several studies show that author contributions are generally unequal and decline with author position [36,36,37,39,40]. The latest review of counting methods in bibliometrics therefore recommends assigning contribution shares based on the order in which authors are listed [55].

However, this pattern has an important exception. Studies of author contributions have shown that first authors usually carry the largest share of the workload [40,42–45], while last authors typically contribute more than middle authors. Informed by this evidence, I employ a modified version of the A-index that assigns the second-largest share to last authors (instead of second authors). Previous studies have shown that this modification improves the validity of counting methods [39,40]. In the multiverse analysis (Section 4.5), I test additional modifications of the A-index that vary the distribution of contribution shares among middle authors. The formulas for all A-index variants tested in this study are provided in the Appendix (Section S3).

## 4.5 Multiverse analysis

This study utilizes multiverse analysis to conduct extensive robustness checks. Instead of asking whether the results hold under one preferred set of assumptions, the multiverse analysis asks whether they remain stable across a range of plausible alternatives [56]. The purpose is to transparently estimate the extent to which research outcomes vary across justifiable modelling assumptions [57].

The multiverse analysis varies four main analytical choices (Table 1). The first choice concerns the distribution of publication credit among middle authors. The sequence-determines-credit model assumes that middle-author order reflects contribution size, so earlier middle authors receive more credit than later middle authors [58]. This is plausible when author lists are intended to reflect descending contribution [39]. The equal contribution model assigns the same credit to all middle authors. This is plausible when middle authors are listed alphabetically or according to conventions unrelated to contribution size, as often occurs in large research teams [41]. The implementation of both models within the mathematical framework of the A-index is detailed in the Appendix (Section S3).

**Table 1: Model specifications that enter the multiverse analysis.**

| Dimension | Specifications | Number of model specifications |
|---|---|---|
| Dataset | 1 – Scopus<br>2 – OpenAlex | 2 |
| Rule for assigning publication credit to middle authors | 1 – Sequence-determines-credit<br>2 – Equal contribution | 2 |
| Minimum percentage of shared DOIs in an author's publication records from Scopus and OpenAlex | 1 – 100% overlap in shared DOIs<br>2 – 90% overlap in shared DOIs | 2 |
| Time window for publication activity | 1 – previous two years<br>2 – previous three years | 2 |
| Total number of model specifications per field | | 16 |

Two further dimensions of the multiverse concern the trade-off between precision and recall in assigning papers to authors. Precision refers to the share of assigned papers that were truly authored by the researcher; recall refers to the share of the researcher's true papers that are captured. I therefore estimate results separately using publication records from OpenAlex and Scopus, which may differ in precision and recall, and vary the required overlap between matched author profiles, using either 100% or 90% shared DOIs. The 100% condition maximizes precision, whereas the 90% condition allows higher recall when one database captures publications missed by the other.

Researchers are defined as active at time $t$ if they published at least one paper between $t$ and $t - x$. In the multiverse analysis, I vary $x$ to test stricter and more lenient definitions of publication activity. Researchers are classified as active if they appear as an author on at least one paper within either the previous two or three years. This involves a trade-off between accuracy and selectivity: a stricter definition more accurately identifies researchers active at the time of measurement but may also make the sample more selective toward researchers who publish frequently.

Out of all possible model specifications within the model space defined in Table 1, I chose a preferred model specification for conducting the main analysis. For the sake of transparency and reproducibility, the main analyses were conducted using OpenAlex data. I require complete DOI overlap with Scopus to minimize the risk of false-positive author-disambiguation errors, that is, papers being assigned to researchers who did not author them. I use the sequence-determines-credit variant of the A-index because equal contribution model tends to be applicable only to very large research teams [41]. Finally, I use a lenient definition of publication activity, in line with previous research [3].

The multiverse analysis tests whether the findings from the preferred model specifications are robust to alternative specifications. The standard indicator of robustness in multiverse analysis is the robustness ratio [59]. The robustness ratio (RR) summarizes both the average estimated trend and its sensitivity to modelling choices. It is defined as the mean coefficient across specifications divided by the total standard error. The total standard error combines two sources of uncertainty: sampling uncertainty within each specification and model uncertainty across specifications. Formally,

$$RR = \frac{\bar{\beta}}{\sqrt{\overline{SE}^2 + SD_{model}^2}}$$

where $\overline{\beta}$ is the mean coefficient across specifications, $\overline{SE}$ is the mean sampling standard error, an $SD_{model}$ is the standard deviation of coefficient estimates across specifications. The robustness ratio is analogous to a $t$-statistic, but it penalizes estimates that vary strongly across plausible analytical choices [59]. A trend is robust if the absolute value of the robustness ratio exceeds 1.96, indicating that the estimate remains statistically distinguishable from zero at the 5% level after accounting for both sampling uncertainty and model uncertainty.

### 4.6 Estimation strategy

This study examines the publication output of early-career researchers using four quantities: (1) the number of published papers; (2) the number of authors per paper; (3) the average individual contribution share, defined as the average A-index value across an individual's papers; and (4) fractional output, defined as the sum of A-index values across an individual's papers. I estimated the number of published papers, and the number of authors per paper using zero-truncated negative binomial regression models, which are well suited to overdispersed count data [60]. Fractionalized publication rates are strictly positive, non-integer values and were therefore modeled using generalized linear models with gamma error distributions and log-link functions. Percent contributions per paper were estimated using a binomial error distribution. In this setting, generalized linear models with log-link functions are preferable to ordinary least squares models with log-transformed outcomes, which can suffer from retransformation bias when predictions are converted back to the original scale [61].

In the main analysis, I used non-parametric specifications with cohort fixed effects rather than imposing a parametric trend across cohorts. This approach makes the central descriptive patterns transparent by avoiding assumptions about the functional form of cross-cohort change. The multiverse analysis uses parametric models to compare the direction and magnitude of estimates across model specifications. All samples in this study are large to calculate sufficiently precise estimates for every field-cohort cell (Table 2).

**Table 2: Number of publication records across model specifications.** The table reports the total number of observations, where each observation is the publication record of a distinct author, for three model specifications: the main model specification, the specification with the smallest analytical sample in the multiverse analysis (S), and the specification with the largest analytical sample in the multiverse analysis (L). It also summarizes the distribution of observations across field-cohort cells, reporting the average, median, minimum, and maximum number of observations per cell.

| Model specification | Total N(obs.) | Mean obs. per cell | Median obs. per cell | Minimum obs. per cell | Maximum obs. per cell |
|---|---|---|---|---|---|
| Main analysis | 90,735 | 907 | 680 | 188 | 2535 |
| Smallest sample | 65,911 | 659 | 492 | 125 | 1912 |
| Largest sample | 211,859 | 2119 | 1454 | 502 | 5148 |

# 5 Results

### 5.1 Main analysis

Figure 1 shows the estimated trends in publication-output metrics from the main model specification. Table 3 displays the total percent changes in these publication output metrics, comparing the earliest and most recent cohorts in the dataset. Because publication and authorship practices differ across disciplines, baseline levels of collaboration and publication output are not directly comparable between fields. This section therefore focuses on changes within each field across cohorts.

Recent cohorts of early-career researchers publish more papers than earlier cohorts within the same career window (Fig. 1a). Across fields, average publication records increased by 22% to 46% (Table 3), corresponding to approximately one to two additional papers per researcher. Because the observation window is fixed, this result indicates increasing publication rates: early-career researchers are publishing more papers over the same period of time.

**Table 3: Total percent changes in outcomes, comparing the 1996-1999 cohorts to the 2011-2014 cohorts.**

| Research field | Papers | Authors per paper | Fractional output |
|---|---|---|---|
| Chem. & materials sci. | 46% | 62% | 8% |
| | (p = 0.000) | (p = 0.000) | (p = 0.019) |
| Earth & env. sci. | 39% | 46% | -4% |
| | (p = 0.000) | (p = 0.000) | (p = 0.156) |
| Engineering | 44% | 47% | -3% |
| | (p = 0.000) | (p = 0.000) | (p = 0.219) |
| Health sci. | 40% | 56% | 1% |
| | (p = 0.000) | (p = 0.000) | (p = 0.574) |
| Life sci. | 22% | 39% | -9% |
| | (p = 0.000) | (p = 0.000) | (p = 0.000) |

Crucially, however, the growth in publication records is matched or exceeded by the growth of collaboration. In all fields, the average number of authors per paper has increased continuously and substantially since 1996 (Fig. 1b). Papers published by the 2015 cohort had, on average, approximately two to three more contributors than papers published by the 1996 cohort. Because early-career researchers share their papers with an increasing number of coauthors, their average contribution share per paper has declined (Fig. 1c).

When publication output is adjusted for collaboration using fractional counting, there is no evidence of a substantial increase in individual publication output across cohorts (Fig. 1d). Trends vary somewhat across fields. In chemistry and materials science, fractional output increased slightly, by approximately 8%, when comparing cohorts from the early 2010s with cohorts from the late 1990s. In all other fields, there is no evidence of increasing fractional output. In the life sciences, collaboration grew more rapidly than publication records, resulting in a small but statistically significant decline in fractional output. In the earth and environmental sciences, health sciences, and engineering, fractional output fluctuates around a field-specific baseline but shows no evidence of systematic change.

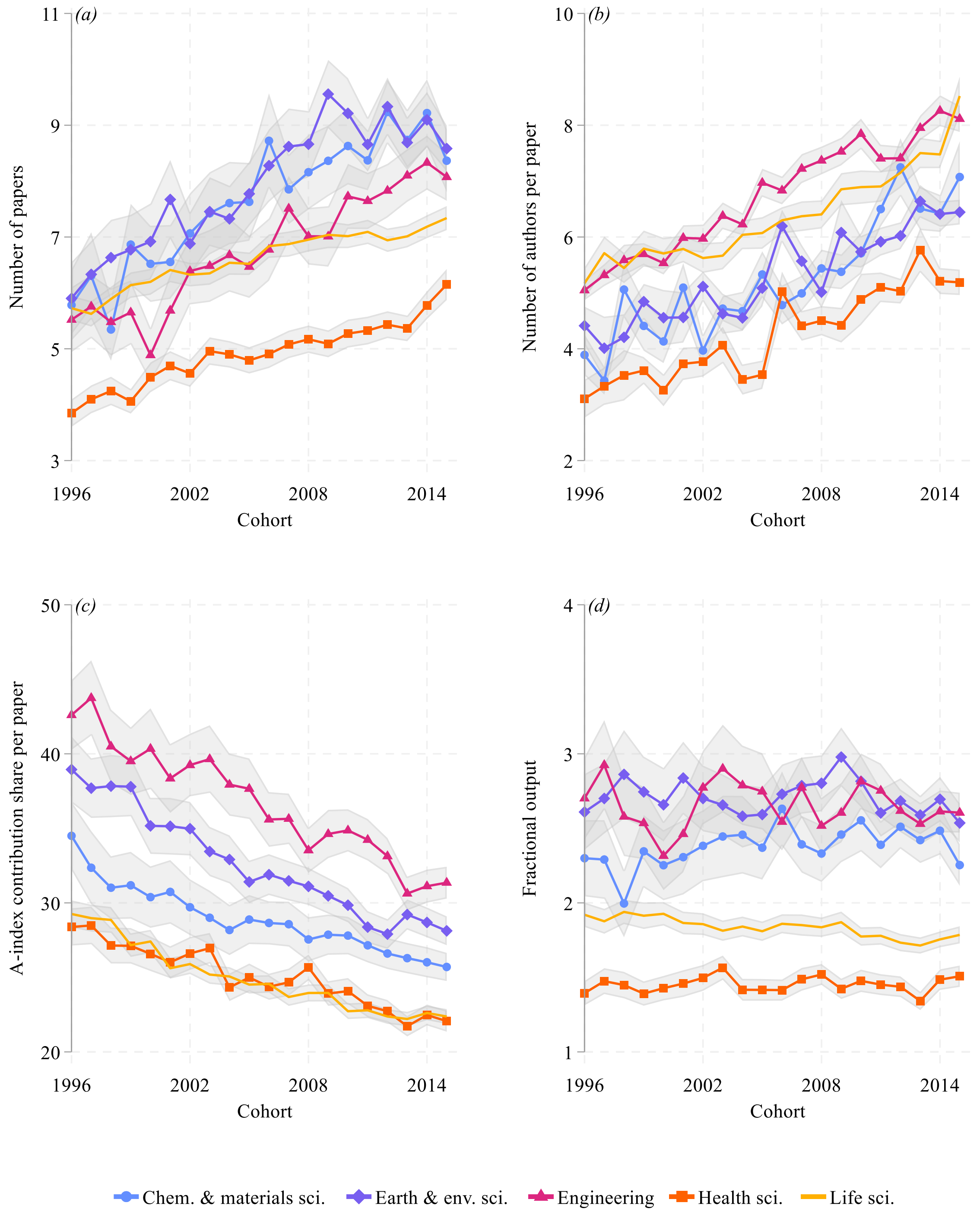

**Figure 1. Trends in publication output among cohorts of early-career researchers, 1996–2015.** a) Estimated number of published papers; b) estimated number of authors per paper; c) estimated contribution share per paper in percent; d) estimated fractional output. All estimates pertain to the first 10 years of publication activity and are plotted by year of the first publication. The plot shows predictive margins and 95% confidence intervals. Number of observations by field: Chem. & materials sci.: N(authors) = 6,564, N(papers) = 97,705. Earth & env. sci.: N(authors) = 11,043, N(papers) = 56,487. Engineering: N(authors) = 8,300, N(papers) = 281,388. Health sci.: N(authors) = 26,885, N(papers) = 64,843. Life sci.: N(authors) = 37,943, N(papers) = 158,699. Papers enter the regressions several times if they appear in the publication records of several authors.

## 5.2 Multiverse analysis

The multiverse analysis confirms that the main conclusion holds across a range of plausible specifications. No field reaches a robustness ratio of 1.96 or higher, indicating that there is no robust evidence of increasing collaboration-adjusted publication output (Table 4). Figure 2 shows the estimated percentage change in fractional output per cohort-year for each field and model specification. Most estimates fall between −0.5% and 0.5%, indicating only small changes over time. Even the most extreme positive and negative estimates suggest that collaboration-adjusted publication output has changed only modestly across cohorts.

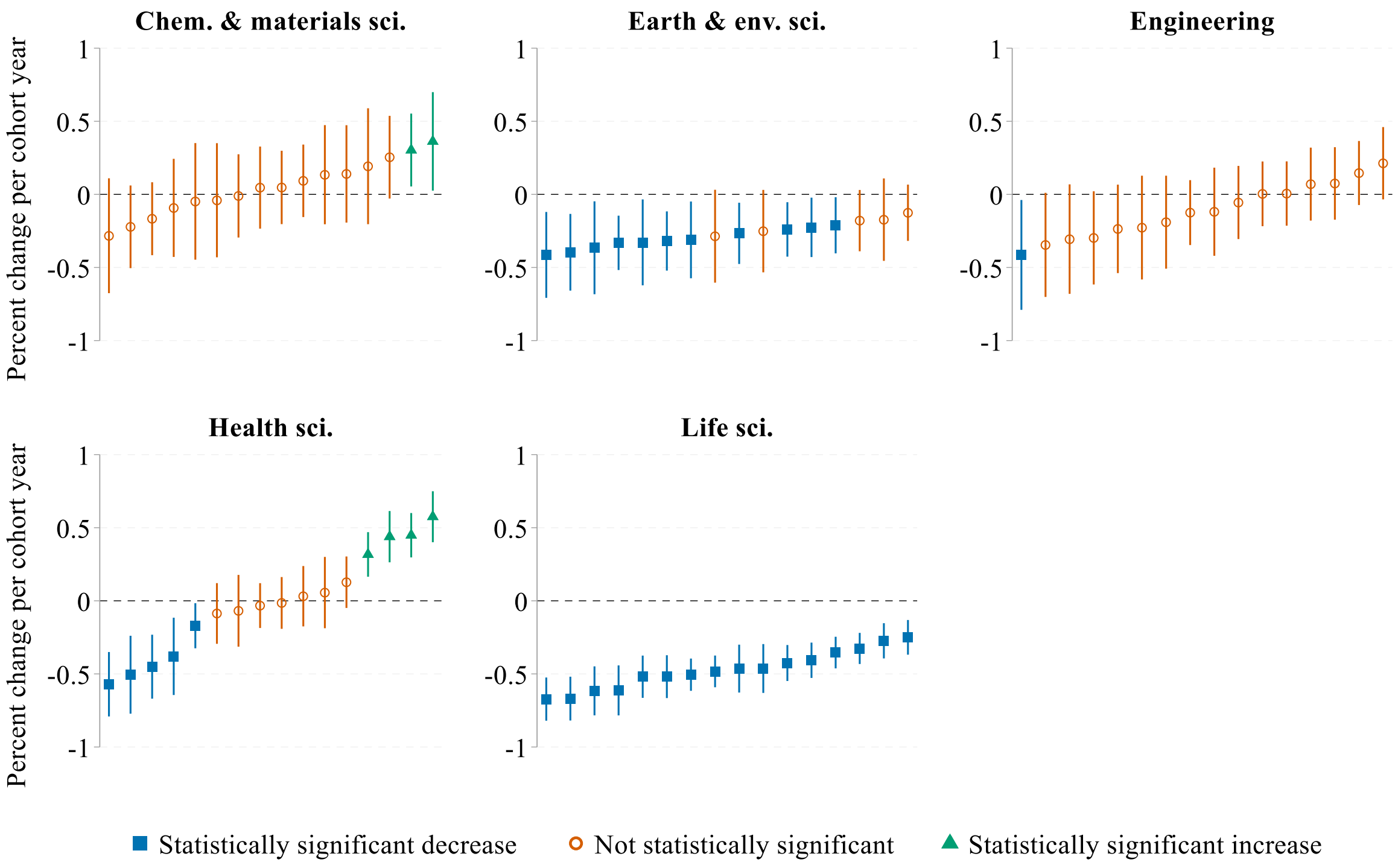

**Figure 2. Estimates and 95% confidence intervals from multiverse analysis by field.** Estimates display the average percent change in fractional output per cohort year and pertain to the 10-year output of cohorts from 1996 through 2020. Estimates are displayed in order from lowest to highest. Depending on the model specification, the total number of observations ranges from N(authors) = 68,292 to N(authors) = 218,913.

**Table 4. Model robustness statistics obtained from multiverse analysis.** Displayed are the average predicted percent change in fractional output per cohort-year, the total standard error, and the robustness ratio. The sign of the robustness ratio indicates the direction of the estimated trend in fractional output across cohorts: negative values indicate decreasing output, and positive values indicate increasing output. Robust trends are defined as |RR| > 1.96. Nonrobust trends, with |RR| ≤ 1.96, should be interpreted as statistically indistinguishable from stable publication rates once both statistical uncertainty and model uncertainty are taken into account.

| Research field | Avg. coefficient | Total SE | Robustness ratio |
|---|---|---|---|
| Chem. & materials sci. | 0.04 | 0.0025 | 0.18 |
| Earth & env. sci. | -0.28 | 0.0015 | -1.83 |
| Engineering | -0.11 | 0.0024 | -0.47 |
| Health sci. | -0.02 | 0.0036 | -0.05 |
| Life sci. | -0.47 | 0.0015 | -3.18 |

In all fields except the health sciences, the estimated trends are largely consistent across analytical choices. For the health sciences, however, results depend primarily on the database used. OpenAlex-based estimates generally show a small increase in fractional output, while Scopus-based estimates show a small decrease. The resulting robustness ratio is very low (-0.05), meaning that the evidence is too fragile to draw a confident conclusion about the direction of the trend.

Overall, the multiverse results are, if anything, less consistent with an increase in fractional output than the main analysis. The modest increase in fractional output observed for chemistry and materials science under the main specification disappears under alternative specifications. In contrast, the life sciences show a highly robust negative trend in the multiverse analysis. The multiverse analysis therefore reinforces the conclusion that fractional output has remained stable or declined across science, engineering, and health fields.

### 5.3 Supporting analyses

The main analysis included only researchers with at least 10 years of publication activity. It therefore intentionally excluded the many early-career researchers who leave academia within a few years of publishing their first paper. A secondary analysis using a much larger sample of researchers with at least 5 years of publication activity provides even stronger evidence of declining fractional output than the main analysis (Appendix, Figure S1, Figure S2, and Table S2).

I also examined larger, less restrictive samples that required less overlap between publication records in Scopus and OpenAlex (Appendix, Figure S3). These samples were not included in the multiverse analysis because specifications with lower overlap thresholds provide less confidence that publication records from the two data providers are accurate. Nonetheless, relaxing the DOI-overlap requirement from 90% to only 75% did not substantially change the research outcome. The resulting trends remained consistent with those observed in the multiverse analysis.

Country-level trends show notable variation (Appendix, Figure S4). In each field, one or two countries deviate from the general pattern. This may partly reflect database coverage: Scopus and OpenAlex primarily index English-language international journals. A country that shifted from non-English regional journals to international outlets may therefore appear to show increasing fractional output. Italy and Spain provide plausible examples. Both increased their presence in international journals after research-assessment reforms [62,63], and their trends differ from those of other countries. Results are more consistent in the United States, the United Kingdom, France, and Germany, where researchers have published mainly in international English-language journals. These countries account for more than 60% of the sample and most analysed papers. No field shows rising fractional output in any of them.

# 6 Discussion

This study found no evidence that the publication counts of early-career researchers at Western research institutions has increased beyond what can be explained by growing collaboration. Although early-career researchers now publish more papers, their average number of coauthors has increased at a similarly high rate. Once publication output is adjusted for collaboration, it has remained remarkably stable since at least the late 1990s. In other words, early-career researchers now have longer publication records than comparable researchers in earlier cohorts because they share more papers with more coauthors. These findings extend and reaffirm earlier research [19].

Although the study uses several supporting analyses to assess the robustness of its findings, its scope is necessarily limited by the nature of publication metadata. Bibliometric data do not directly observe individual contributions to papers; contributions must instead be inferred from authorship. Authorship is an imperfect proxy because some contributors may not be listed as authors, a practice known as ghost authorship, while some listed authors may have made little or no meaningful contribution, a practice described as honorary or gift authorship. For such practices to affect the study's conclusions, authorship patterns would need to have changed substantially over time. To my knowledge, there is no evidence that this has occurred [64]. Moreover, some evidence suggests that authorship standards for early-career researchers may be relatively lenient, such that they receive authorship credit even for comparatively small contributions [65]. Overall, the use of authorship as a proxy for contribution is unlikely to explain this study's finding of stable collaboration-adjusted output.

More generally, bibliometric data cannot measure how much time and effort early-career researchers invest in the papers they publish. Research may have become more complex, increasing the total effort required to produce a publishable study [23]. The growing number of authors per paper is consistent with this interpretation: it seems unlikely that researchers would routinely distribute an unchanged amount of work among an increasingly large number of collaborators. Rather, larger research teams may reflect the greater amount or specialization of work required for a single publication. It is not clear whether collaboration reduces or increases the overall workload. Dividing tasks among team members may make the research process more efficient. At the same time, collaboration may come with inherent costs: larger teams can create additional demands for coordination, conflict resolution, and project management. These considerations highlight that publication output cannot be equated with workload. Although the fractional publication output of early-career researchers has remained stable, the amount of work required to produce that output may have changed. An important task for future research is to determine whether the time and effort required to produce a publishable paper have changed over time.

The findings are limited to researchers in selected science, engineering, and health fields at Western research institutions. Publication rates may have developed differently in other fields and regions. Future research should therefore build on this study by examining populations not covered here, including early-career researchers at non-Western research institutions and in fields with different publication cultures. China's research output is difficult to study because of author-disambiguation challenges. However, it would be an important case for future research, given its high publication output [66], intense pressure to publish [25,67,68], and large research teams [69].

# 7 Conclusion

The meaning of publication counts has changed as research has become more collaborative over time. A single paper increasingly represents the distributed effort of many contributors rather than the concentrated effort of one or a few researchers. This has important implications for research evaluation: raw publication counts are no longer an adequate measure of individual publication output. Crucially, this problem also affects commonly used bibliometric indicators such as the h-index [70] as well as journal-based indicators used in evaluation contexts, which generally do not account for collaboration. Researchers who contribute smaller shares to many coauthored papers may accumulate publications and citations at a higher rate than researchers who concentrate their efforts on fewer contributions. Evaluation systems that rely heavily on unadjusted metrics may therefore reward participation in large collaborative publication networks rather than individual research efforts. A practical response should be to reduce the role of raw publication counts in evaluation. In an increasingly collaborative research system, fair evaluation requires asking not only how many papers a researcher has published, but what role they played in producing them.

For related reasons, the growing number of papers listed on early-career researchers' CVs cannot, by itself, be treated as evidence that they increasingly use questionable practices to inflate publication counts. As this study shows, the increase in publication records can be explained by growing collaboration. This does not rule out the possibility that salami-slicing and similar practices have become more common. However, it severely weakens the argument that the publication patterns of contemporary researchers provide evidence for such a trend. To maintain that argument, one would need to show that researchers increasingly engage in salami slicing even though their fractional output has not increased.

More broadly, these findings point to the need for a better understanding of how researchers navigate potential conflicts between publication incentives and research integrity. Researchers should not be viewed as passive recipients of incentives. Most (early-career) researchers likely have strong reasons to care about the quality and integrity of their work, and compromising these standards may be a last resort rather than the default response to publication pressure. The relevant risk factor may therefore not be publication pressure alone, but whether researchers have the resources to manage it without compromising their work. Where researchers have sufficient time, stable resources, collaboration networks, and appropriate training, publication pressure need not translate into poor research practices [71,72]. Where these supports are absent, the risk of questionable research practices is likely much greater. Future research should therefore examine how working conditions and institutional support shape how researchers respond to publication incentives.

# Ethical Statement

This work did not require ethical approval from a human subject or animal welfare committee.

# Data accessibility

This manuscript is currently under review at a peer-reviewed journal. The data and code will be made available upon publication of the final version of the study.

# Declaration of AI use

OpenAI GPT-5.6 Terra was used for final language editing after the manuscript's conceptual and argumentative development had been completed. Anthropic Claude Sonnet 5 was used to assist with coding in Stata and PostgreSQL, as well as with generating comments for the replication package. The author reviewed and edited all resulting content as needed and takes full responsibility for the content of the publication.

# Authors' contribution

C.L.: conceptualization, data curation, formal analysis, investigation, methodology, software, validation, visualization, writing – original draft, writing – review and editing

# Competing interests

I have no competing interests.

# Funding statement

This study utilizes bibliometric metadata provided by the German Kompetenznetzwerk Bibliometrie, which has been funded since December 2008 by the Federal Ministry of Research, Technology and Space of Germany (BMFTR) under grant numbers 16WIK2101A, 16WIK2101B, and 16WIK2101C.

# Acknowledgements

Early versions of this project were presented at the seminar "Analytical Sociology: Theory and Empirical Applications" at Venice International University in November 2025 and at the 30th Annual International Conference on Science and Technology Indicators at the University of Antwerp in September 2026. I thank the participants for their constructive feedback. An early version of this study was published in the conference proceedings of the 30th Annual International Conference on Science and Technology Indicators [73].

I thank my colleagues in the Department of Sociology at LMU Munich for their valuable feedback: Agnieszka Althaber, Alisia Bauer, Christian Ganser, Fabian Kalleitner, Laura Schächtele, Renata Topinkova, Felix Wolter, and Anna Zamberlan. I also thank Katrin Auspurg (LMU Munich), Lutz Bornmann (Administrative Headquarters of the Max Planck Society), and Tobias Wolbring (Friedrich-Alexander University Erlangen-Nürnberg) for their helpful feedback on the research and manuscript. I am especially grateful to Robin Haunschild (Max Planck Institute for Solid State Research) for suggesting the integration of publication records from Scopus and OpenAlex.

# References

1. Cyranoski D, Gilbert N, Ledford H, Nayar A, Yahia M. 2011 Education: The PhD factory. *Nature* **472**, 276–279. (doi:10.1038/472276a)
2. Warren JR. 2019 How much do you have to publish to get a job in a top sociology department? Or to get tenure? Trends over a generation. *Sociol. Sci.* **6**, 172–196. (doi:10.15195/v6.a7)
3. Milojević S, Radicchi F, Walsh JP. 2018 Changing demographics of scientific careers: The rise of the temporary workforce. *Proc. Natl. Acad. Sci. U. S. A.* **115**, 12616–12623. (doi:10.1073/pnas.1800478115)
4. Haubrock PJ *et al.* 2026 How contemporary academic structures constrain scientific creativity and hold back early-career researchers. *Nat. Hum. Behav.* Advance online publication. (doi:10.1038/s41562-026-02575-5)
5. Tijdink JK, Vergouwen ACM, Smulders YM. 2013 Publication pressure and burn out among Dutch medical professors: A nationwide survey. *PloS One* **8**, e73381. (doi:10.1371/journal.pone.0073381)
6. Dreisoerner A, Goetz V, Frohnmayer D, Tran US, Voracek M, Nater UM. 2026 Prevalence and severity of mental health problems in early-career researchers: A systematic review and meta-analysis. *Nat. Hum. Behav.* **10**, 1698–1717. (doi:10.1038/s41562-026-02505-5)
7. Johann D, Raabe IJ, Rauhut H. 2022 Under pressure: The extent and distribution of perceived pressure among scientists in Germany, Austria, and Switzerland. *Res. Eval.* **31**, 385–409. (doi:10.1093/reseval/rvac014)
8. Johann D, Neufeld J, Thomas K, Rathmann J, Rauhut H. 2024 The impact of researchers' perceived pressure on their publication strategies. *Res. Eval.* Advance online publication. (doi:10.1093/reseval/rvae011)
9. Tiokhin L, Yan MH, Morgan TJH. 2021 Competition for priority harms the reliability of science, but reforms can help. *Nat. Hum. Behav.* **5**, 857–867. (doi:10.1038/s41562-020-01040-1)
10. Sarewitz D. 2016 The pressure to publish pushes down quality. *Nature* **533**, 147–147. (doi:10.1038/533147a)
11. Wright DE. 2024 Five problems plaguing publishing in the life sciences - and one common cause. *FEBS Lett.* **598**, 2227–2239. (doi:10.1002/1873-3468.15018)
12. Smaldino PE, McElreath R. 2016 The natural selection of bad science. *R. Soc. Open Sci.* **3**, 160384. (doi:10.1098/rsos.160384)
13. Kohrt F, Smaldino PE, McElreath R, Schönbrodt F. 2023 Replication of the natural selection of bad science. *R. Soc. Open Sci.* **10**, 221306. (doi:10.1098/rsos.221306)
14. Brischoux F, Angelier F. 2015 Academia's never-ending selection for productivity. *Scientometrics* **103**, 333–336. (doi:10.1007/s11192-015-1534-5)
15. Bauldry S. 2013 Trends in the research productivity of newly hired assistant professors at research departments from 2007 to 2012. *Am. Sociol.* **44**, 282–291. (doi:10.1007/s12108-013-9187-4)
16. Wuchty S, Jones BF, Uzzi B. 2007 The increasing dominance of teams in production of knowledge. *Science* **316**, 1036–1039. (doi:10.1126/science.1136099)
17. Huang DW. 2015 Temporal evolution of multi-author papers in basic sciences from 1960 to 2010. *Scientometrics* **105**, 2137–2147. (doi:10.1007/s11192-015-1760-x)
18. Nabout JC, Parreira MR, Teresa FB, Carneiro FM, da Cunha HF, Ondei LD, Caramori SS, Soares TN. 2015 Publish (in a group) or perish (alone): The trend from single- to multi-authorship in biological papers. *Scientometrics* **102**, 357–364. (doi:10.1007/s11192-014-1385-5)
19. Fanelli D, Larivière V. 2016 Researchers' individual publication rate has not increased in a century. *PloS One* **11**, e0149504. (doi:10.1371/journal.pone.0149504)
20. Parish AJ, Boyack KW, Ioannidis JPA. 2018 Dynamics of co-authorship and productivity across different fields of scientific research. *PloS One* **13**, e0189742. (doi:10.1371/journal.pone.0189742)
21. Fortunato S *et al.* 2018 Science of science. *Science* **359**, eaao0185. (doi:10.1126/science.aao0185)
22. Stoltz DS. 2023 Rising coauthorship in sociology, 1895 to 2022. *Socius* **9**, 23780231231171115. (doi:10.1177/23780231231171115)

23. Cordero RJB, de León-Rodriguez CM, Alvarado-Torres JK, Rodriguez AR, Casadevall A. 2016 Life science's average publishable unit (APU) has increased over the past two decades. *PloS One* **11**, e0156983. (doi:10.1371/journal.pone.0156983)
24. Gandevia S. 2018 Publication pressure and scientific misconduct: Why we need more open governance. *Spinal Cord* **56**, 821–822. (doi:10.1038/s41393-018-0193-9)
25. Qiu J. 2010 Publish or perish in China. *Nature* **463**, 142. (doi:10.1038/463142a)
26. Gopalakrishna G, Ter Riet G, Vink G, Stoop I, Wicherts JM, Bouter LM. 2022 Prevalence of questionable research practices, research misconduct and their potential explanatory factors: A survey among academic researchers in The Netherlands. *PloS One* **17**, e0263023. (doi:10.1371/journal.pone.0263023)
27. Schneider JW, Allum N, Andersen JP, Petersen MB, Madsen EB, Mejlgaard N, Zachariae R. 2024 Is something rotten in the state of Denmark? Cross-national evidence for widespread involvement but not systematic use of questionable research practices across all fields of research. *PloS One* **19**, e0304342. (doi:10.1371/journal.pone.0304342)
28. Tijdink JK, Schipper K, Bouter LM, Maclaine Pont P, de Jonge J, Smulders YM. 2016 How do scientists perceive the current publication culture? A qualitative focus group interview study among Dutch biomedical researchers. *BMJ Open* **6**, e008681. (doi:10.1136/bmjopen-2015-008681)
29. Schneck A. 2023 Are most published research findings false? Trends in statistical power, publication selection bias, and the false discovery rate in psychology (1975–2017). *PloS One* **18**, e0292717. (doi:10.1371/journal.pone.0292717)
30. Vankov I, Bowers J, Munafò MR. 2014 On the persistence of low power in psychological science. *Q. J. Exp. Psychol.* **67**, 1037–1040. (doi:10.1080/17470218.2014.885986)
31. Hicks D. 2012 Performance-based university research funding systems. *Res. Policy* **41**, 251–261. (doi:10.1016/j.respol.2011.09.007)
32. Frietsch R, Gruber S, Neuhäusler P. 2026 Erfassung und Analyse bibliometrischer Indikatoren 2026 im Rahmen des Pakt-Monitorings zum Pakt für Forschung und Innovation IV. See https://www.bmftr.bund.de/SharedDocs/Downloads/DE/2026/bibliometriebericht2026.pdf?__blob=publicationFile&v=1 (accessed on 17 September 2026).
33. Bornmann L, Haunschild R, Mutz R. 2021 Growth rates of modern science: A latent piecewise growth curve approach to model publication numbers from established and new literature databases. *Humanit. Soc. Sci. Commun.* **8**, 224. (doi:10.1057/s41599-021-00903-w)
34. Horbach S. 2020 Pandemic publishing: Medical journals strongly speed up their publication process for COVID-19. *Quant. Sci. Stud.* **1**, 1056–1067. (doi:10.1162/qss_a_00076)
35. Haupka N, Culbert JH, Schniedermann A, Jahn N, Mayr P. 2026 Analysis of the publication and document types in OpenAlex, Web of Science, Scopus, PubMed and Semantic Scholar. *Quant. Sci. Stud.* **7**, 179–194. (doi:10.1162/QSS.a.406)
36. Donner P. 2020 A validation of coauthorship credit models with empirical data from the contributions of PhD candidates. *Quant. Sci. Stud.* **1**, 551–564. (doi:10.1162/qss_a_00048)
37. Donner P, Vincent-Lamarre P. 2025 Almost always unequal: Co-authors' contributions to scientific publications. In *Proceedings of the 20th International Conference on Scientometrics & Informetrics* (eds Shushanik Sargsyan, Wolfgang Glänzel, Giovanni Abramo), pp. 223–236. Yerevan, Armenia. (doi:10.51408/issi2025_017)
38. Ali JR. 2021 Quantitative author inputs to Earth science research publications: Survey results, insights and potential applications. *Geol. Mag.* **158**, 951–963. (doi:10.1017/s0016756820000916)
39. Hagen NT. 2010 Harmonic publication and citation counting: Sharing authorship credit equitably - not equally, geometrically or arithmetically. *Scientometrics* **84**, 785–793. (doi:10.1007/s11192-009-0129-4)
40. Sundling P. 2023 Author contributions and allocation of authorship credit: Testing the validity of different counting methods in the field of chemical biology. *Scientometrics* **128**, 2737–2762. (doi:10.1007/s11192-023-04680-y)
41. Mongeon P, Smith E, Joyal B, Larivière V. 2017 The rise of the middle author: Investigating collaboration and division of labor in biomedical research using partial alphabetical authorship. *PloS One* **12**, e0184601. (doi:10.1371/journal.pone.0184601)
42. Larivière V, Desrochers N, Macaluso B, Mongeon P, Paul-Hus A, Sugimoto CR. 2016 Contributorship and division of labor in knowledge production. *Soc. Stud. Sci.* **46**, 417–435. (doi:10.1177/0306312716650046)

43. Larivière V, Pontille D, Sugimoto CR. 2021 Investigating the division of scientific labor using the Contributor Roles Taxonomy (CRediT). *Quant. Sci. Stud.* **2**, 111–128. (doi:10.1162/qss_a_00097)
44. Sauermann H, Haeussler C. 2017 Authorship and contribution disclosures. *Sci. Adv.* **3**, e1700404. (doi:10.1126/sciadv.1700404)
45. Corrêa Jr. EA, Silva FN, da F. Costa L, Amancio DR. 2017 Patterns of authors contribution in scientific manuscripts. *J. Informetr.* **11**, 498–510. (doi:10.1016/j.joi.2017.03.003)
46. Schmidt M, Rimmert C, Stephen D, Lenke C, Donner P, Gärtner S, Taubert N, Bausenwein T, Stahlschmidt S. 2025 The data infrastructure of the German Kompetenznetzwerk Bibliometrie: An enabling intermediary between raw data and analysis. *Quant. Sci. Stud.* **6**, 1129–1146. (doi:10.1162/QSS.a.20)
47. Aman V. 2018 Does the Scopus author ID suffice to track scientific international mobility? A case study based on Leibniz laureates. *Scientometrics* **117**, 705–720. (doi:10.1007/s11192-018-2895-3)
48. Andersen JP. 2023 Field-level differences in paper and author characteristics across all fields of science in Web of Science, 2000–2020. *Quant. Sci. Stud.* **4**, 394–422. (doi:10.1162/qss_a_00246)
49. Nane GF, Larivière V, Costas R. 2017 Predicting the age of researchers using bibliometric data. *J. Informetr.* **11**, 713–729. (doi:10.1016/j.joi.2017.05.002)
50. Dinh L, Barley WC, Johnson L, Allan BF. 2024 Hyperauthored papers disproportionately amplify important egocentric network metrics. *Quant. Sci. Stud.* **5**, 613–636. (doi:10.1162/qss_a_00307)
51. Sánchez-Jiménez R, Rodriguez JT, Guerrero-Bote VP, de Moya-Anegón F. 2026 Identification of hyperauthorship in scientific publications: An approach based on outlier detection methods. *J. Informetr.* **20**, 101803. (doi:10.1016/j.joi.2026.101803)
52. Nielsen MW, Andersen JP. 2021 Global citation inequality is on the rise. *Proc. Natl. Acad. Sci. U. S. A.* **118**, e2012208118. (doi:10.1073/pnas.2012208118)
53. Savage WE, Olejniczak AJ. 2022 More journal articles and fewer books: Publication practices in the social sciences in the 2010's. *PloS One* **17**, e0263410. (doi:10.1371/journal.pone.0263410)
54. Stallings J, Vance E, Yang JS, Vannier MW, Liang JM, Pang LJ, Dai L, Ye I, Wang G. 2013 Determining scientific impact using a collaboration index. *Proc. Natl. Acad. Sci. U. S. A.* **110**, 9680–9685. (doi:10.1073/pnas.1220184110)
55. Gauffriau M. 2021 Counting methods introduced into the bibliometric research literature 1970-2018: A review. *Quant. Sci. Stud.* **2**, 932–975. (doi:10.1162/qss_a_00141)
56. Steegen S, Tuerlinckx F, Gelman A, Vanpaemel W. 2016 Increasing transparency through a multiverse analysis. *Perspect. Psychol. Sci.* **11**, 702–712. (doi:10.1177/1745691616658637)
57. Leibel C, Bornmann L. 2026 Introducing multiverse analysis to bibliometrics: The case of team size effects on disruptive research. *Quant. Sci. Stud.* **7**, 695–716. (doi:10.1162/QSS.a.475)
58. Tscharntke T, Hochberg ME, Rand TA, Resh VH, Krauss J. 2007 Author sequence and credit for contributions in multiauthored publications. *PLoS Biol.* **5**, 13–14. (doi:10.1371/journal.pbio.0050018)
59. Young C, Holsteen K. 2017 Model uncertainty and robustness: A computational framework for multimodel analysis. *Sociol. Methods Res.* **46**, 3–40. (doi:10.1177/0049124115610347)
60. Hilbe JM. 2014 *Modeling count data*. Cambridge: Cambridge University Press. (doi:10.1017/CBO9781139236065)
61. Duan N. 1983 Smearing estimate - A nonparametric retransformation method. *J. Am. Stat. Assoc.* **78**, 605–610. (doi:10.2307/2288126)
62. Jiménez-Contreras E, Anegón FD, López-Cózar ED. 2003 The evolution of research activity in Spain: The impact of the National Commission for the Evaluation of Research Activity (CNEAI). *Res. Policy* **32**, 123–142. (doi:10.1016/s0048-7333(02)00008-2)
63. Abramo G, D'Angelo CA. 2021 The different responses of universities to introduction of performance-based research funding. *Res. Eval.* **30**, 514–528. (doi:10.1093/reseval/rvab022)
64. Wislar JS, Flanagin A, Fontanarosa PB, DeAngelis CD. 2011 Honorary and ghost authorship in high impact biomedical journals: A cross sectional survey. *Br. Med. J.* **343**, d6128. (doi:10.1136/bmj.d6128)
65. Pruschak G, Hopp C. 2022 And the credit goes to ... - Ghost and honorary authorship among social scientists. *PloS One* **17**, e0267312. (doi:10.1371/journal.pone.0267312)
66. Ioannidis JPA, Collins TA, Baas J. 2024 Evolving patterns of extreme publishing behavior across science. *Scientometrics* **129**, 5783–5796. (doi:10.1007/s11192-024-05117-w)

67. Hvistendahl M. 2013 China's publication bazaar. *Science* **342**, 1035–1039. (doi:10.1126/science.342.6162.1035)
68. Horta H, Li H. 2023 Nothing but publishing: The overriding goal of PhD students in mainland China, Hong Kong, and Macau. *Stud. High. Educ.* **48**, 263–282. (doi:10.1080/03075079.2022.2131764)
69. Liu LL, Yu JF, Huang JM, Xia F, Jia T. 2021 The dominance of big teams in China's scientific output. *Quant. Sci. Stud.* **2**, 350–362. (doi:10.1162/qss_a_00099)
70. Hirsch JE. 2005 An index to quantify an individual's scientific research output. *Proc. Natl. Acad. Sci. U. S. A.* **102**, 16569–16572. (doi:10.1073/pnas.0507655102)
71. Yeo-Teh NSL, Tang BL. 2022 Perceived publication pressure and research misconduct: Should we be too bothered with a causal relationship? *Res. Ethics* **18**, 329–338. (doi:10.1177/17470161221125097)
72. Fanelli D, Costas R, Larivière V. 2015 Misconduct policies, academic culture and career stage, not gender or pressures to publish, affect scientific integrity. *PloS One* **10**, e0127556. (doi:10.1371/journal.pone.0127556)
73. Leibel C. 2026 Rising publication counts do not imply rising individual output among early-career researchers. In *30th Annual International Conference on Science and Technology Indicators. Conference Proceedings* (eds Ine De Parade, Eline Vandewalle, Tim C.E. Engels), pp. 661–674. Antwerp: ECOOM. (doi:10.5281/zenodo.22656518)
74. Caron E, Eck NJ van. 2014 Large scale author name disambiguation using rule-based scoring and clustering. In *Proceedings of the Science and Technology Indicators conference* (ed Ed Noyons), pp. 79–86. Leiden, Netherlands.
75. Tekles A, Bornmann L. 2020 Author name disambiguation of bibliometric data: A comparison of several unsupervised approaches. *Quant. Sci. Stud.* **1**, 1510–1528. (doi:10.1162/qss_a_00081)

# Appendix

# Supporting Information Text

### S1. Alternative Author Disambiguation Methods

I considered three alternative approaches to constructing individual publication records. The first was to rely exclusively on ORCID records. However, ORCID is based on voluntary registration and is therefore unsuitable as the sole basis for historical cohort analyses. This is especially problematic because selection into ORCID differs across cohorts. Among recent cohorts, ORCID possession is less selective because many journals now require or strongly encourage ORCID registration. Among older cohorts, by contrast, researchers could create an ORCID profile only if they remained active after ORCID was introduced in 2012. Even among those researchers, ORCID users are likely to represent a selective group of early adopters whose publication trajectories may differ from those of the average researcher.

The second option was to sample researchers with highly distinctive names, such as authors with three surnames. This strategy was used by prior research on the publication rates of early-career researchers [19]. This method can reduce name ambiguity, but it introduces other limitations. It may oversample researchers from regions where such naming conventions are common, and it yields a comparatively small sample. Such a sample is poorly suited to the present estimation strategy, which requires large numbers of observations to estimate cohort trends across fields and subgroups.

The third option was to use an algorithmic method for constructing individual publication records [74] that performed well in a validation study [75]. However, this method can ensure high accuracy only for researchers with a minimum level of publication output. Prior research has applied this algorithm only to authors with at least five papers [52]. This restriction is unsuitable for studying trends in the publication output of early-career researchers, because researchers with low publication output are an important part of the target population.

The triangulation approach used in this study addresses these limitations. Unlike ORCID-based records, OpenAlex and Scopus author identifiers are algorithmically generated and therefore do not depend on researchers opting into a profile system. Unlike distinctive-name sampling, the approach yields a large sample without explicitly selecting researchers from regions or naming traditions where highly distinctive names are more common. Finally, unlike algorithmic methods that require a minimum number of publications, the comparison of shared DOIs in Scopus and OpenAlex provides a way to identify high-confidence publication records even for researchers with few publications. The approach therefore combines scale, reduced selection bias, and measurement confidence for the population most relevant to this study.

### S2. Supplementary Information on Data Processing

Following prior research [19], I assigned researchers to countries and fields using majority rule. Researchers linked to multiple countries through their affiliations were assigned to the country appearing most often in their publication record; researchers linked to multiple fields were assigned to the field appearing most often in their publications. Researchers were placed in an “unknown” category in the case of ties.

To determine researchers’ primary country of activity, I relied primarily on affiliation data from Scopus. This choice reflects a current limitation of OpenAlex: affiliation information is often missing or incomplete [50,51]. OpenAlex affiliation data were used only when Scopus data were insufficient to assign a primary country. Combining publication records across databases allows the analysis to draw on the relative strengths of each data source when constructing researcher-level variables.

Since the target population consists of early-career researchers in academia, researchers were retained in the analytical sample only if they were affiliated with an academic research institution at the time of measurement. Academic affiliations were identified using a combination of the institution's name and, where available, its entry in the Research Organization Registry (ROR). To reduce errors caused by missing affiliation data in OpenAlex, researchers were retained if they were affiliated with an academic institution in either Scopus or OpenAlex.

To ensure consistent field assignment across datasets, I used the field-level of the OpenAlex topics classification (Appendix, Table S1). Based on recommendations in prior research, I grouped fields with similar bibliometric characteristics, such as comparable team sizes and publication practices [48]. Research fields were grouped into intentionally broad categories: chemistry and materials sciences, earth and environmental sciences, engineering, health sciences, and life sciences. Broad categories improve statistical power, especially in subgroup analyses, and, more importantly, allows most researchers to be assigned to a field by majority rule. A more fine-grained field classification would make it harder to assign researchers with only a few publications to a field. For this reason, broad field categories are better suited to studying trends in publication output than a highly detailed field classification.

### S3. A-index variants

This study uses an adapted version of the A-index to account for the special role of last authors in the fields analyzed. I therefore assign the largest credit share to the first author and the second-largest credit share to the last author. Middle authors are treated in two alternative ways. The first version follows the sequence-determines-credit (SDC) model. This model assumes that the order of middle authors reflects contribution size. For papers with fewer than three authors, the standard A-index is used:

$$A_{i,n}^{SDC} = E\left(x_{i(n)}\right) \; for \; n < 3$$

For papers with three or more authors, the adapted SDC version is defined as:

$$A_{i,n}^{SDC} = \begin{cases} E\left(x_{1(n)}\right), \; i = 1 \\ E\left(x_{2(n)}\right), \; i = n \\ E\left(x_{(i+1)(n)}\right), \; 1 < i < n \end{cases} \quad for \; n \geq 3$$

Thus, the first author receives the highest credit share, the last author receives the credit share normally assigned to the second author, and all middle authors are moved one position down in the credit ranking.
The second version follows the equal-contribution (EC) model for middle authors. This model assumes that middle-author order does not reliably encode differences in contribution. Again, for papers with fewer than three authors, the standard A-index is used:

$$A_{i,n}^{EC} = E\left(x_{i(n)}\right) \; for \; n < 3$$

For papers with three or more authors, the adapted EC version is defined as:

$$A_{i,n}^{EC} = \begin{cases} E\left(x_{1(n)}\right), \; i = 1 \\ E\left(x_{2(n)}\right), \; i = n \\ \dfrac{1 - E\left(x_{1(n)}\right) - E\left(x_{2(n)}\right)}{n - 2}, \; 1 < i < n \end{cases} \quad for \; n \geq 3$$

In this version, the first and last authors receive the same credit as in the adapted SDC model, but all middle authors divide the remaining credit equally. The SDC and EC versions therefore differ only in how they allocate credit among middle authors. This distinction is used in the multiverse analysis to test whether the results depend on assumptions about the contribution order of middle authors.

# Figures

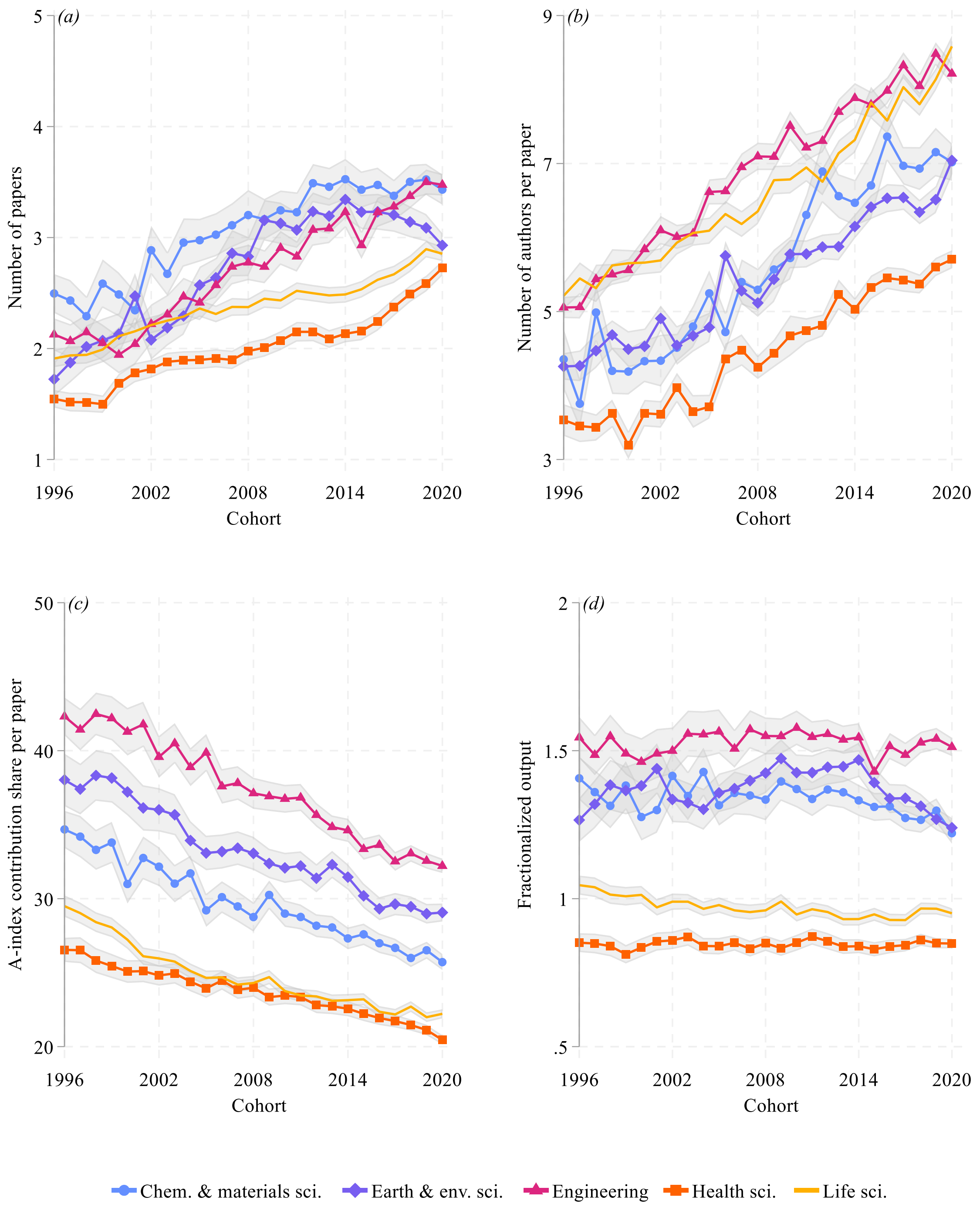

**Figure S1. Trends in publication output among cohorts of early-career researchers, 1996–2020.** a) Estimated number of published papers; b) Estimated number of papers published as first author; c) Estimated contribution share per paper in percent; d) Estimated fractionalized output. All outcomes pertain to the first 5 years of publication activity and are plotted by year of first publication. The plot shows predictive margins and 95% confidence intervals. Number of observations by field: Chem. & materials sci.: N(authors) = 30,558, N(papers) = 211,237. Earth & env. sci.: N(authors) = 32,905, N(papers) = 185,264. Engineering: N(authors) = 42,894, N(papers) = 639,155. Health sci.: N(authors) = 97,809, N(papers) = 244,349. Life sci.: N(authors) = 114,465, N(papers) = 463,481. Papers enter the regressions several times if they appear in the publication records of several authors.

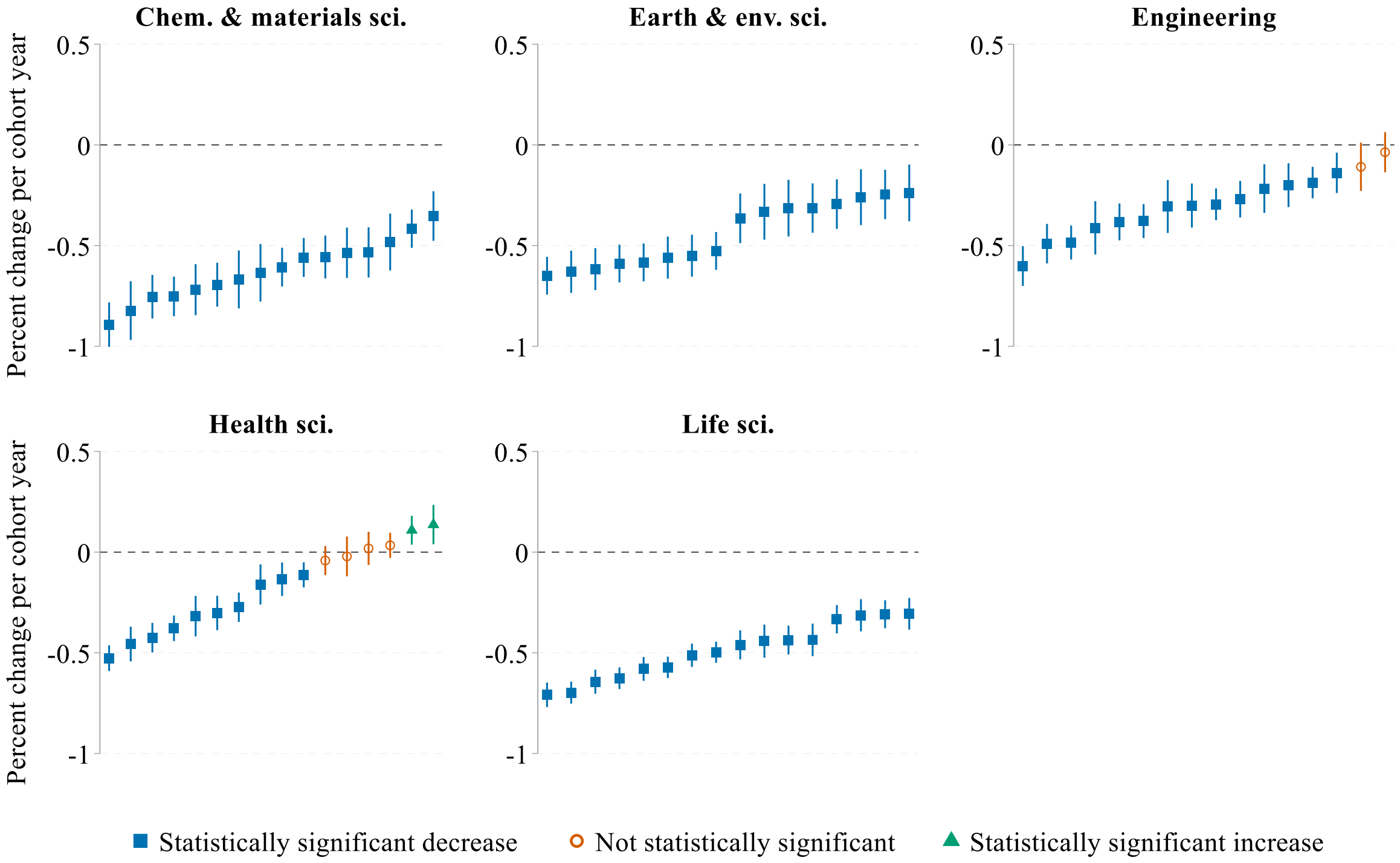

**Figure S2. Estimates and 95% confidence intervals from multiverse analysis by field.** Estimates display the average percent change in fractional output per cohort year and pertain to the 5-year output of cohorts from 1996 through 2020. Estimates are displayed in order from lowest to highest. Depending on the model specification, the total number of observations ranges from N = 227,174 to N = 579,350.

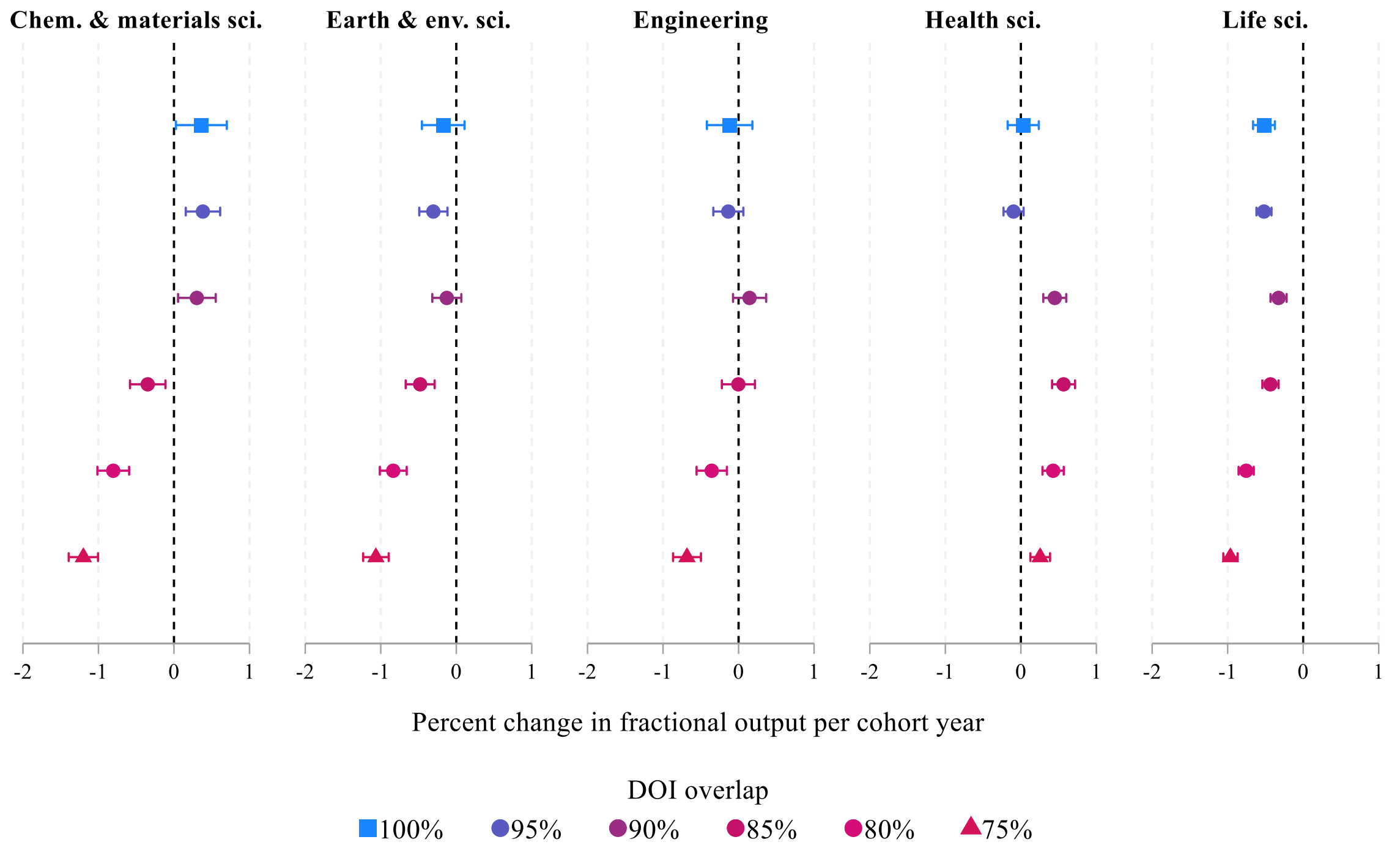

**Figure S3. Estimates and 95% confidence intervals of cross-cohort growth rates in fractional output.** The estimates originate from different analytical samples, which were constructed using several thresholds for the minimum required overlap of DOIs in publication records form OpenAlex and Scopus. The higher the threshold, the more the independent disambiguation algorithms of OpenAlex and Scopus agree that a specific individual published a specific set of papers. All estimates pertain to the 10-year output of cohorts from 1996 through 2015. The total number of observations ranges from N = 90,735 to N = 288,273.

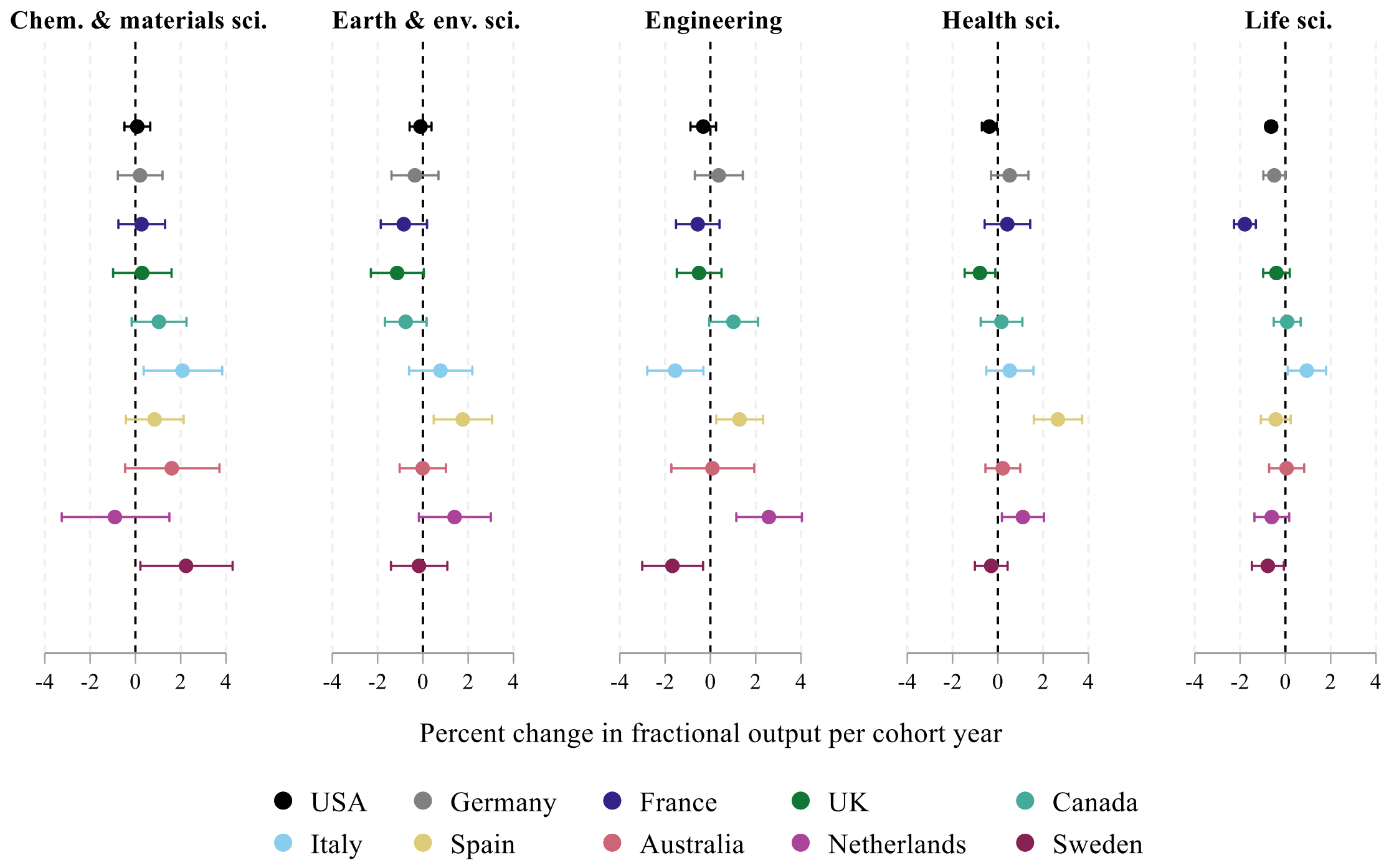

**Figure S4. Estimates and 95% confidence intervals of cross-cohort growth rates in the fractionalized publication output of early-career researchers by country.** Countries are ordered by their share of authors in the dataset, based on primary institutional affiliation. The ten countries shown account for about 88% of all early-career researchers in the analytical sample used in the main analysis (10-year observation period). Researchers primarily affiliated with institutions in the USA account for about 37% of early-career researchers in the analytical sample. Number of observations by country: USA: N = 33,417. Germany: N = 9,142. France: N = 6,494. UK: N= 6,353. Canada: N = 5,724. Italy: N = 4,769. Spain: N = 4,586. Australia: N = 3,829. The Netherlands: N = 2,910. Sweden: N = 2,635.

# Tables

**Table S1.** Classification of OpenAlex field-level topics.

| OpenAlex field | Classification |
|---|---|
| Chemistry; Materials science | Chemistry & materials sciences |
| Earth and planetary sciences; Environmental science | Earth & environmental sciences |
| Chemical engineering; Energy; Engineering | Engineering |
| Agricultural and biological sciences; Biochemistry, genetics and molecular biology; Immunology and microbiology; Neuroscience | Life sciences |
| Dentistry; Health professions; Pharmacology, toxicology and pharmaceutics; Medicine; Nursing; Veterinary | Health sciences |

**Table S2.** Model robustness statistics obtained from multiverse analysis. Estimates display the average percent change in fractional output per cohort year and pertain to the 5-year output of cohorts from 1996 through 2020. Displayed are the average predicted percent change in fractional output per cohort-year, the total standard error, and the robustness ratio.

| **Research field** | **Avg. coefficient** | **Total SE** | **Robustness ratio** |
|---|---|---|---|
| Chem. & materials sci. | -0.62 | 0.0016 | -3.91 |
| Earth & env. sci. | -0.44 | 0.0017 | -2.63 |
| Engineering | -0.30 | 0.0016 | -1.86 |
| Health sci. | -0.18 | 0.0021 | -0.84 |
| Life sci. | -0.49 | 0.0014 | -3.49 |